\documentclass[twocolumn,superscriptaddress,nofootinbib,floatfix,aps,prb,citeautoscript,longbibliography]{revtex4-2}
\usepackage[utf8]{inputenc}
\usepackage[T1]{fontenc}
\usepackage{lineno}

\usepackage{graphicx}
\usepackage{color}
\usepackage{array}
\usepackage{dcolumn}
\usepackage{bm}
\usepackage{multirow}
\usepackage[version=4]{mhchem}
\usepackage{cleveref}
\usepackage{tabularx}
\usepackage[usenames,dvipsnames]{xcolor} 
\usepackage{siunitx}
\usepackage{comment}
\usepackage{tikz}
\usepackage{ulem}

\newcommand{\Tc}{\ensuremath{T_\textrm{c}}}

\newcommand{\olog}{\ensuremath{\omega_\text{log}}}
\newcommand{\dosef}{\ensuremath{N_\text{F}}}

\newcommand{\aF}{\ensuremath{\alpha^2F}}

\begin{document}

\newcommand{\bochum}{Research Center Future Energy Materials and Systems of the University Alliance Ruhr and Interdisciplinary Centre for Advanced Materials Simulation, Ruhr University Bochum, Universitätsstraße 150, D-44801 Bochum, Germany}
\newcommand{\coimbra}{CFisUC, Department of Physics, University of Coimbra, Rua Larga, 3004-516 Coimbra, Portugal}
\newcommand{\mpi}{Max-Planck-Institut f\"ur Mikrostrukturphysik, Weinberg 2, D-06120 Halle, Germany}
\newcommand{\upv}{Fisika Aplikatua Saila, Gipuzkofako Ingeniaritza Eskola, University of the Basque Country (UPV/EHU), Europa Plaza 1, 20018 Donostia/San Sebastián, Spain}
\newcommand{\cfismat}{Centro de Física de Materiales (CSIC-UPV/EHU), Manuel de Lardizabal Pasealekua 5, 20018 Donostia/San Sebastián, Spain}
\newcommand{\dipc}{Donostia International Physics Center (DIPC), Manuel de Lardizabal Pasealekua 4, 20018 Donostia/San Sebastián, Spain}
\newcommand{\ludwig}{Chair for Materials Discovery and Interfaces, Institute for Materials, Ruhr University Bochum, Universitätsstraße 150, D-44801 Bochum, Germany}

\author{Antonio Sanna}
\affiliation{\mpi}
\author{Tiago F. T. Cerqueira}
\affiliation{\coimbra}
\author{Yue-Wen Fang}
\affiliation{\upv}
\affiliation{\cfismat}
\author{Ion Errea}
\affiliation{\upv}
\affiliation{\cfismat}
\affiliation{\dipc}
\author{Alfred Ludwig}
\affiliation{\ludwig}
\author{Miguel A. L. Marques} 
\email{miguel.marques@rub.de}
\affiliation{\bochum} 

\date{\today}

\title{Prediction of Ambient Pressure Conventional Superconductivity above 80~K in Thermodynamically Stable Hydride Compounds}

\begin{abstract}
The primary challenge in the field of high-temperature superconductivity in hydrides is to achieve a superconducting state at ambient pressure rather than the extreme pressures that have been required in experiments so far. Here, we propose a family of compounds, of composition \ce{Mg2XH6} with X$=$Rh, Ir, Pd, or Pt, that achieves this goal. These materials were identified by scrutinizing more than a million compounds using a machine-learning accelerated high-throughput workflow. They are thermodynamically stable, indicating that they are serious candidates for experimental synthesis. We predict that their superconducting transition temperatures are in the range of 45--80~K, or even above 100~K with appropriate electron doping of the Pt compound. These results indicate that, although very rare, high-temperature superconductivity in thermodynamically stable hydrides is achievable at room pressure. 
\end{abstract}

\maketitle

\section{Introduction}

Superconductivity was first discovered by Kamerlingh Onnes in 1911 when he observed mercury's electrical resistance drop to zero at extremely low temperatures~\cite{onnes1911superconductivity}. 
Despite the discovery of many new superconducting metals in the following decades, 
all of them required temperatures near absolute zero to exhibit their superconducting properties. A major breakthrough came in 1986 when Müller and Bednorz discovered high-temperature superconductivity, above the boiling point of nitrogen, in a class of cuprate compounds~\cite{Bednorz1986,Bednorz1988_585}.

Superconducting materials are usually divided into conventional and unconventional depending on the physical mechanism that binds the Cooper pairs. In contrast to unconventional superconductivity, where an elusive electronic mechanism is at play, in  conventional superconductors the interaction between the electrons at the Fermi level and the vibrations of the lattice is responsible for the pairing. In this case, the transition temperature is related directly to the average frequency of the relevant phonons and to the strength of their coupling to the electrons.
Naturally, a straightforward way to maximize the phonon frequencies is to reduce the mass of the constituent chemical elements.  This led Ashcroft to propose that atomic metallic hydrogen had the potential to exhibit superconductivity at remarkably high temperatures, even at or near room temperature~\cite{ashcroft1968metallicH-PRL,wigner1935possibility}.  Unfortunately, molecular hydrogen only dissociates and forms an atomic metal at very high pressures, larger than the ones that can be currently reached in high-pressure experiments~\cite{DalladaySimpson2016,eremets_semimetallic_2019,loubeyre_synchrotron_2020,monacelli_quantum_2023}.

To circumvent this problem, researchers turned to hydride compounds~\cite{FloresLivasReveiwHydrides2020}, where one can achieve metallicity at considerably lower pressures, due to the effect of chemical pre-compression of the hydrogen~\cite{Ashcroft2004,Hilleke2022,Gilman_1971_PRL_LiHF}. The first hydride that was found to superconduct at high temperature was \ce{SH3} at pressures exceeding 100~GPa~\cite{Ma_SH3_JChemPhys2014,Duan_SH3_SciRep2014,Drozdov2015}, marking the first demonstration of high-temperature superconductivity in hydrides. Researchers continued to explore various hydrogen-rich hydrides under extreme pressures, searching for new materials with even higher critical temperatures for superconductivity. These efforts have led to the discovery of several high-temperature superconductors such as LaH$_{10}$ ($T_c$ of 250~K at 170~GPa)~\cite{Hemley_LaH_PRL2019,Eremets_LaH_Nature2019,Errea2020-LaH10Nature} and CaH$_6$ ($T_c$ of 215~K at 172~GPa)~\cite{CaH6-RRL-YanmingMA2022} among others.

Currently, the primary challenge in the field of superconductivity in hydrides is to achieve superconductivity at lower, or ideally, at ambient pressure rather than the extreme pressures that have been required in experiments so far~\cite{Boeri_SCRoadmap2021,FloresLivasReveiwHydrides2020,Ferreira_LuNHsystem_NatComm2023,Boeri_Boron_PRB2020}. Unfortunately, this is far from trivial. On the one hand, hydrides at low pressure are typically insulating and therefore not superconducting. On the other hand, most superconducting systems at high-pressure become dynamically unstable upon pressure release, even if ionic quantum fluctuations and the consequent anharmonicity can dynamically stabilize high-$T_c$ compounds at much lower pressures than expected when these effects are neglected~\cite{Errea2020-LaH10Nature,Errea_Quantum_2016}.

Few hydrides have been suggested in the literature to superconduct at high temperature and ambient pressure. For example, \ce{Al4H} in space group $Pm\Bar{3}m$, inspired by the perovskite structure with Al in both the $1b$ and $3d$ position with H in the $1a$ position, was predicted to have a transition temperature of up to 54~K~\cite{He2023a}. In addition \ce{(Be4)2H}, composed of Be hexagonal layers intercalated with hexagonal layers of H, was predicted to have a \Tc\ in the range of 72--84~K~\cite{He2023b}. Unfortunately, these materials are \textit{thermodynamically} unstable, meaning that it is energetically favorable for them to decompose into other compounds, making very difficult their experimental synthesis and characterization. In fact, \ce{Al4H} is at 136~meV/atom above the convex hull of thermodynamic stability, decomposing to \ce{AlH3} and  Al, while \ce{(Be4)2H} is 202~meV/atom above the hull, decomposing to \ce{BeH2} and elemental Be.

\begin{figure}[htb]
  \centering
    \includegraphics[width=5cm]{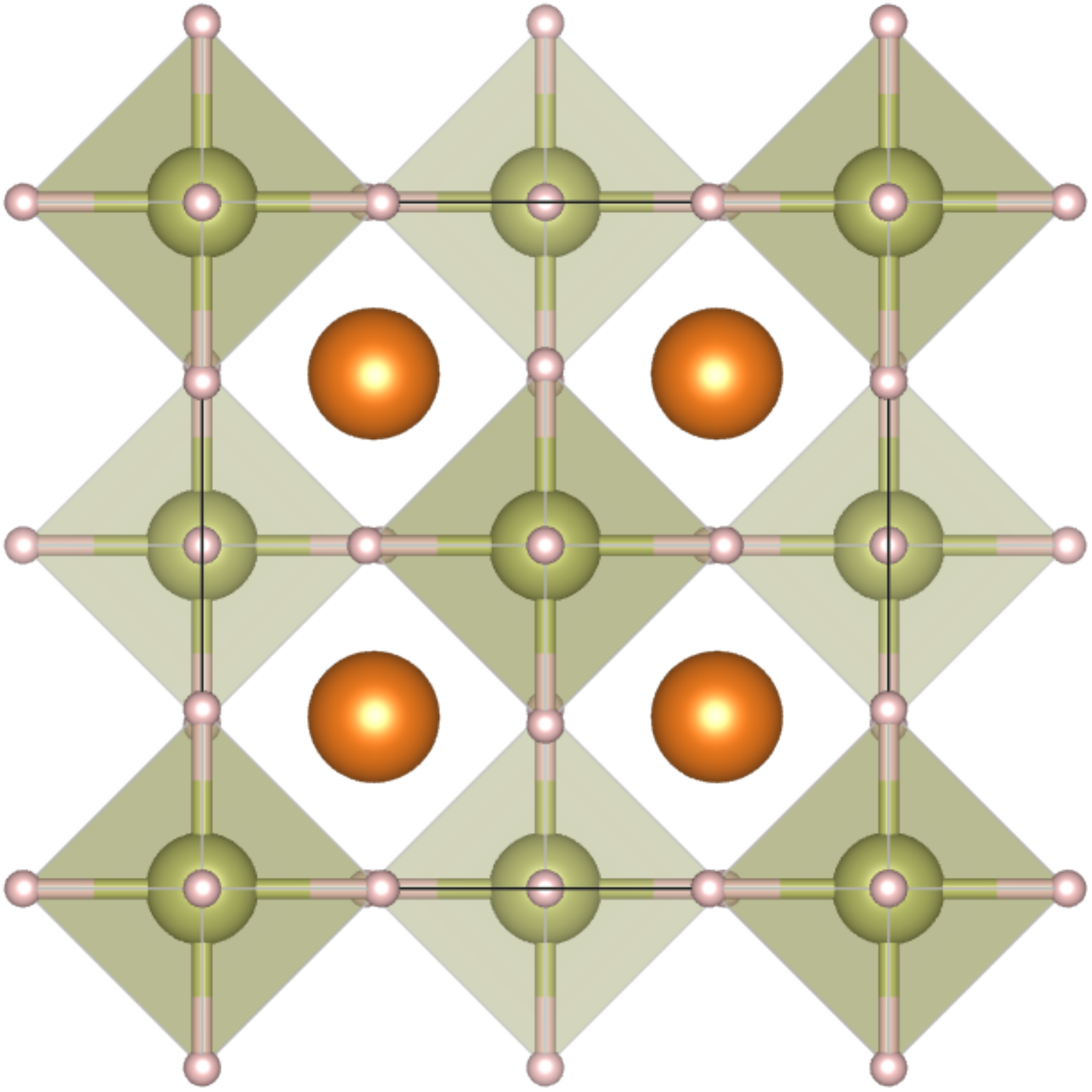}
    \caption{The face-centered cubic conventional unit cell of \ce{Mg2XH6}. The X atoms are in green, Mg atoms in orange, and H atoms in white.}
  \label{fig:crystalstructure}
\end{figure}

Here we present a family of metallic compounds that we predict to be high-temperature conventional superconductors, but that are also thermodynamically stable. These materials have chemical composition \ce{Mg2XH6} where X is Rh, Ir, Pd, or Pt, and are related to \ce{Mg2RuH6}, a compound that has been previously synthesized experimentally~\cite{Kritikos1991,Huang1991}. To identify these compounds, we used a machine-learning accelerated search~\cite{newAdvMat} of more than 1 million compounds present in the Alexandria database~\cite{SciAdv,AdvMat}.

\section{Results and discussion}

The conventional unit cell of \ce{Mg2XH6} is depicted in \cref{fig:crystalstructure}. In this structure with space group $Fm\Bar{3}m$ the X atoms are in the 4a Wyckoff positions, the Mg atoms in the 8c, and the H atoms in the 24e positions (optimized structures available in the SI). Note that the H atoms form octahedra around the X atoms (with a X--H distance of 1.70~\AA, 1.72~\AA, 1.78~\AA, and 1.79~\AA\ for X$=$Rh, Ir, Pd, and Pt respectively), but the octahedra do not share any vertices, edges, or faces as it is common in perovskite systems. We note that this is the same structure type found experimentally for \ce{Mg2RuH6}~\cite{Kritikos1991,Huang1991}. The main difference is that when the transition metal is Ru, belonging to group VIII of the periodic table, the compound becomes a semiconductor with a rather large band-gap. For X$=$Rh, Ir, Pd, and Pt, which belong to groups IX and X of the periodic table, the band structure around the Fermi-energy is very similar, showing the electronic similarity between all these compounds, but the extra electrons dope the system and place the Fermi level in the conduction band.

\definecolor{stablecolor}{RGB}{50,162,162}
\definecolor{linecolor}{RGB}{75,200,200}
\definecolor{unstablecolor}{RGB}{200,75,200}
\definecolor{gray}{RGB}{240,240,240}
\newcommand\bolden[1]{{\boldmath\bfseries#1}}

\begin{figure}[htb]
  \begin{tikzpicture}[x=6cm,y=6cm]
\draw[gray] (0.9, 0.0) -- (0.45, 0.7794228634059948);
\draw[gray] (0.1, 0.0) -- (0.55, 0.7794228634059948);
\draw[gray] (0.05, 0.08660254037844387) -- (0.9500000000000001, 0.08660254037844387);
\draw[gray] (0.8, 0.0) -- (0.4, 0.6928203230275509);
\draw[gray] (0.2, 0.0) -- (0.6000000000000001, 0.6928203230275509);
\draw[gray] (0.1, 0.17320508075688773) -- (0.9, 0.17320508075688773);
\draw[gray] (0.7, 0.0) -- (0.35, 0.606217782649107);
\draw[gray] (0.30000000000000004, 0.0) -- (0.65, 0.606217782649107);
\draw[gray] (0.15000000000000002, 0.2598076211353316) -- (0.85, 0.2598076211353316);
\draw[gray] (0.6, 0.0) -- (0.3, 0.5196152422706631);
\draw[gray] (0.4, 0.0) -- (0.7, 0.5196152422706631);
\draw[gray] (0.2, 0.34641016151377546) -- (0.8, 0.34641016151377546);
\draw[gray] (0.5, 0.0) -- (0.25, 0.4330127018922193);
\draw[gray] (0.5, 0.0) -- (0.75, 0.4330127018922193);
\draw[gray] (0.25, 0.4330127018922193) -- (0.75, 0.4330127018922193);
\draw[gray] (0.4, 0.0) -- (0.2, 0.34641016151377546);
\draw[gray] (0.6, 0.0) -- (0.8, 0.34641016151377546);
\draw[gray] (0.3, 0.5196152422706631) -- (0.7, 0.5196152422706631);
\draw[gray] (0.29999999999999993, 0.0) -- (0.14999999999999997, 0.2598076211353315);
\draw[gray] (0.7000000000000001, 0.0) -- (0.8500000000000001, 0.2598076211353315);
\draw[gray] (0.35000000000000003, 0.6062177826491071) -- (0.6499999999999999, 0.6062177826491071);
\draw[gray] (0.19999999999999996, 0.0) -- (0.09999999999999998, 0.17320508075688767);
\draw[gray] (0.8, 0.0) -- (0.9, 0.17320508075688767);
\draw[gray] (0.4, 0.6928203230275509) -- (0.6, 0.6928203230275509);
\draw[gray] (0.09999999999999998, 0.0) -- (0.04999999999999999, 0.08660254037844384);
\draw[gray] (0.9, 0.0) -- (0.95, 0.08660254037844384);
\draw[gray] (0.45, 0.7794228634059948) -- (0.55, 0.7794228634059948);
\draw[gray] (0.0, 0.0) -- (0.0, 0.0);
\draw[gray] (1.0, 0.0) -- (1.0, 0.0);
\draw[gray] (0.5, 0.8660254037844386) -- (0.5, 0.8660254037844386);\input{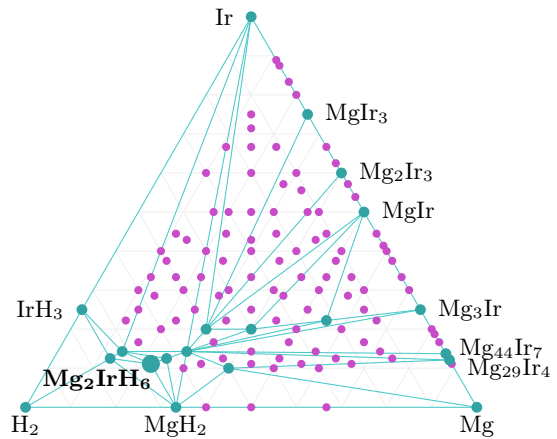}\end{tikzpicture}
  \caption{Ternary phase diagram of the Mg--Ir--H compositions studied in this work computed with PBEsol. Blue circles denote thermodynamically stable phases, while small magenta circles denote low-energy metastable phases that are within 100~meV/atom from the convex hull. The high-temperature superconducting phase is indicated in bold.}
  \label{fig:pd_comp}
\end{figure}

From the thermodynamic point of view, the compounds with X$=$Rh, Ir, and Pt are all on the convex hull of stability, while \ce{Mg2PdH6} is 56~meV/atom above it, with the most favorable decomposition channel being \ce{MgPdH4} and \ce{MgH2}. The formation energies are $-720$~meV/atom, $-730$~meV/atom, $-519$~meV/atom, and $-563$~meV/atom for X$=$Rh, Ir, Pd, Pt respectively, with a stability that seems to increase slightly with the period.

To understand better the phase diagram of these hydrides, we plot in \cref{fig:pd_comp} the theoretical phase diagram of Mg--X--Ir. This diagram was constructed considering experimental phases, machine-learning accelerated prototype search~\cite{SciAdv,AdvMat}, and structures obtained with systematic global structure prediction runs~\cite{Goedecker2004,Amsler2010} (see sec.~\ref{sec:methods}). We see that the phase diagram includes a series of thermodynamically stable phases, specifically \ce{Mg3Ir2H5}, \ce{Mg2IrH2}, \ce{MgIrH5}, \ce{Mg5Ir2H2}, \ce{Mg2IrH4}, \ce{Mg2IrH5}, \ce{MgIrH6}, \ce{Mg2IrH6}, and \ce{Mg4IrH5}. Also a very large number of phases are meta-stable at relatively small distance to the convex hull. 

A possible way to identify these predicted compounds is by applying high-throughput experimentation using thin film materials libraries and their high-throughput characterization~\cite{Ludwig2019}. Indeed, the search for superconductors is a classic example of combinatorial materials science~\cite{Xiang1995} and the metal to hydride transformation and the associated change of optical and mechanical properties was already used for searching hydrogen storage materials~\cite{Gremaud2007, Ludwig2007}.

Thin films offer high-purity synthesis across various chemistries in a clean vacuum environment, with most elements available as source materials (e.g., sputter targets). They are prepared either by depositing metallic films exposed to hydrogen at high pressures and temperatures, or by direct synthesis in a hydrogen-containing atmosphere. Reactive sputtering in an Ar/\ce{H2} mixture facilitates direct hydride formation, even for composition spread thin films~\cite{Mongstad2012}. Post-deposition annealing in high-pressure pure \ce{H2} and high temperatures supports hydride phase formation.
Predicted compounds with a well-defined stoichiometry can be efficiently sought using composition spread libraries. The predicted phase can then be confirmed by comparing calculated XRD patterns to high-throughput measurements. Superconducting properties of the identified phase are subsequently determined through electrical four-point probe and cryogenic magnetic measurements.

\begin{figure*}[htb]
  \centering
  \includegraphics[height=10cm]{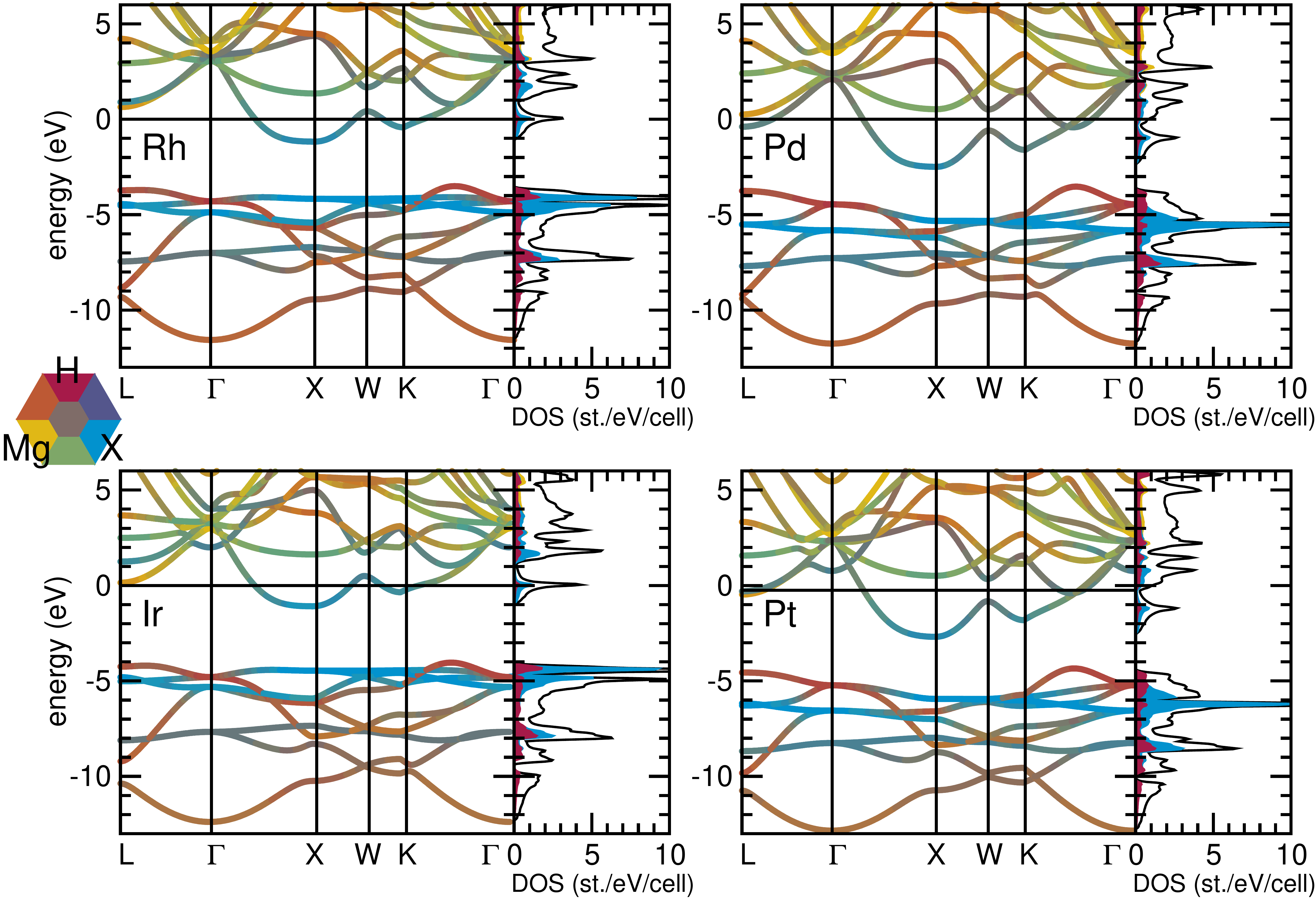}
  \caption{Electronic band structures and density of states of \ce{Mg2RhH6} (top left), \ce{Mg2IrH6} (bottom left), \ce{Mg2PdH6} (top right), and \ce{Mg2PtH6} (bottom right). The color-scale reports the projection of the Kohn-Sham states into atomic sites.}
  \label{fig:elbands}
\end{figure*}

\begin{figure*}[htb]
  \centering
  \includegraphics[height=10cm]{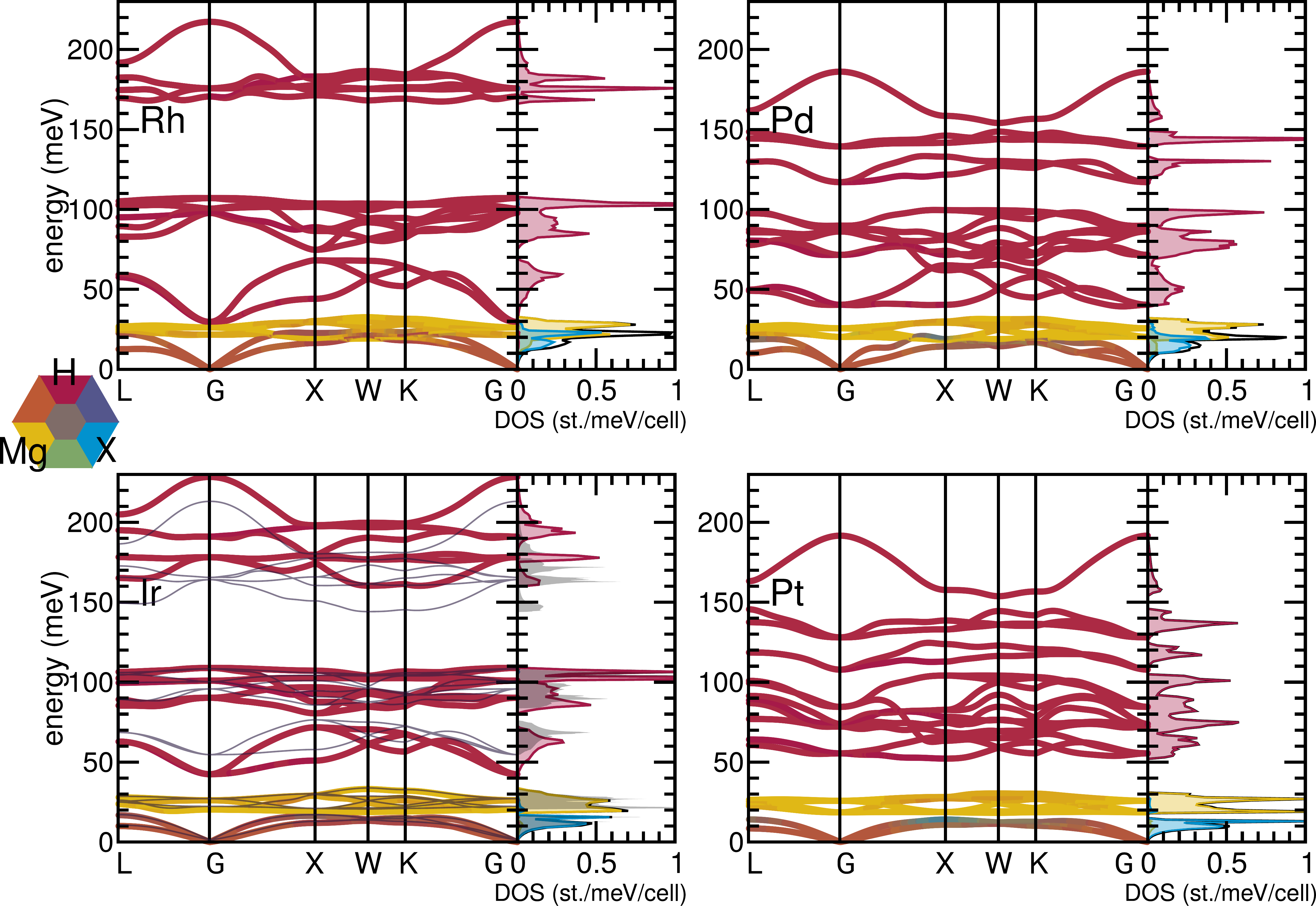}
  \caption{Phonon band structures and density of states of \ce{Mg2RhH6} (top left), \ce{Mg2IrH6} (bottom left), \ce{Mg2PdH6} (top right), and \ce{Mg2PtH6} (bottom right).  The color-scale reports the projection of the atomic displacements into atomic sites. The thin gray line in the Ir panel is the anharmonic phonon band structure as simulated with the SSCHA method.}
  \label{fig:phbands}
\end{figure*}

In \cref{fig:elbands} we present the electronic band structure for \ce{Mg2RhH6}, \ce{Mg2IrH6}, \ce{Mg2PdH6}, and \ce{Mg2PtH6}. All these materials are metallic but with a very peculiar electronic structure in the vicinity of the Fermi level, characterized by two major van Hove singularities, a peak and a valley. The Rh and Ir compounds have the Fermi level pinned to a peak in the density of states, while the Pd and Pt compounds have it on a shoulder slightly above a minimum due to their extra electron. The density of states at the Fermi level is significantly larger in the Ir and Rh compounds, however it decreases very rapidly (within the phononic energy scale) moving away from the Fermi level. 

A second crucial aspect to observe is that the character of the Fermi level states, while dominated by the X atoms, shows a significant H content, which is a crucial ingredient needed to reach a high \Tc\ in hydrides~\cite{belli2021strong}.
This aspect is very uncommon at room pressure, because H states tend to be either fully occupied and deep below the Fermi level or fully unoccupied and far above the Fermi level. Charge projections indicate a fraction of H charge of the order of 1/5 of the total Fermi level charge. 
The H contribution appears to be distributed evenly among the Fermi level states.

The calculated vibrational spectrum in the harmonic approximation is reported in \cref{fig:phbands}. The phonon band structure is decomposed according to the phonon displacement of the mode, while in the density of states we consider the eigenfunctions of the dynamical matrix. This allows to identify better the atomic component of the mode.
The most important aspect is that the phonon modes are stable in all four compounds. The vibrational spectrum is similar for all materials with three well identifiable regions. At low energy, from 0 to 20~meV, we have the acoustic and the X-atom modes (owing to the high mass of X ions). In the 20--30~meV range we find the Mg modes. 
Above 30~meV and up to 230~meV we observe the H modes. In the Rh and Ir compounds, hydrogen modes are well separated into a middle energy region between 30 and 100~meV and a high energy window between 160 and 230 meV. In the Pd and Pt compounds they have a more compact spectrum spanning from 40 to 190~meV. 

In many hydrides the lattice dynamics are strongly affected by anharmonicity~\cite{Errea2013,Errea_Quantum_2016,Errea2020-LaH10Nature}, deeply impacting the predicted superconducting critical temperature. To verify if this is the case for our systems, we performed anharmomic phonon calculations for the Ir compound with the stochastic self-
consistent harmonic approximation (SSCHA)~\cite{Errea-PRB2014-SSCHA,Bianco-PRB2017-SSCHA,monacelli2018pressure,Monacelli2021}, which is presently the gold standard for anharmonic lattice dynamic simulations in solids. We observe that anharmonicity has some effect on the phonon spectra, slightly softening the highest energy H-character modes and hardening the lowest energy H-character modes. Due to the similarity of the harmonic phonon spectra in all compounds, a similar weak impact of anharmonicity is expected for all of them.

\begin{figure}[htb]
  \centering
  \includegraphics[width=8cm]{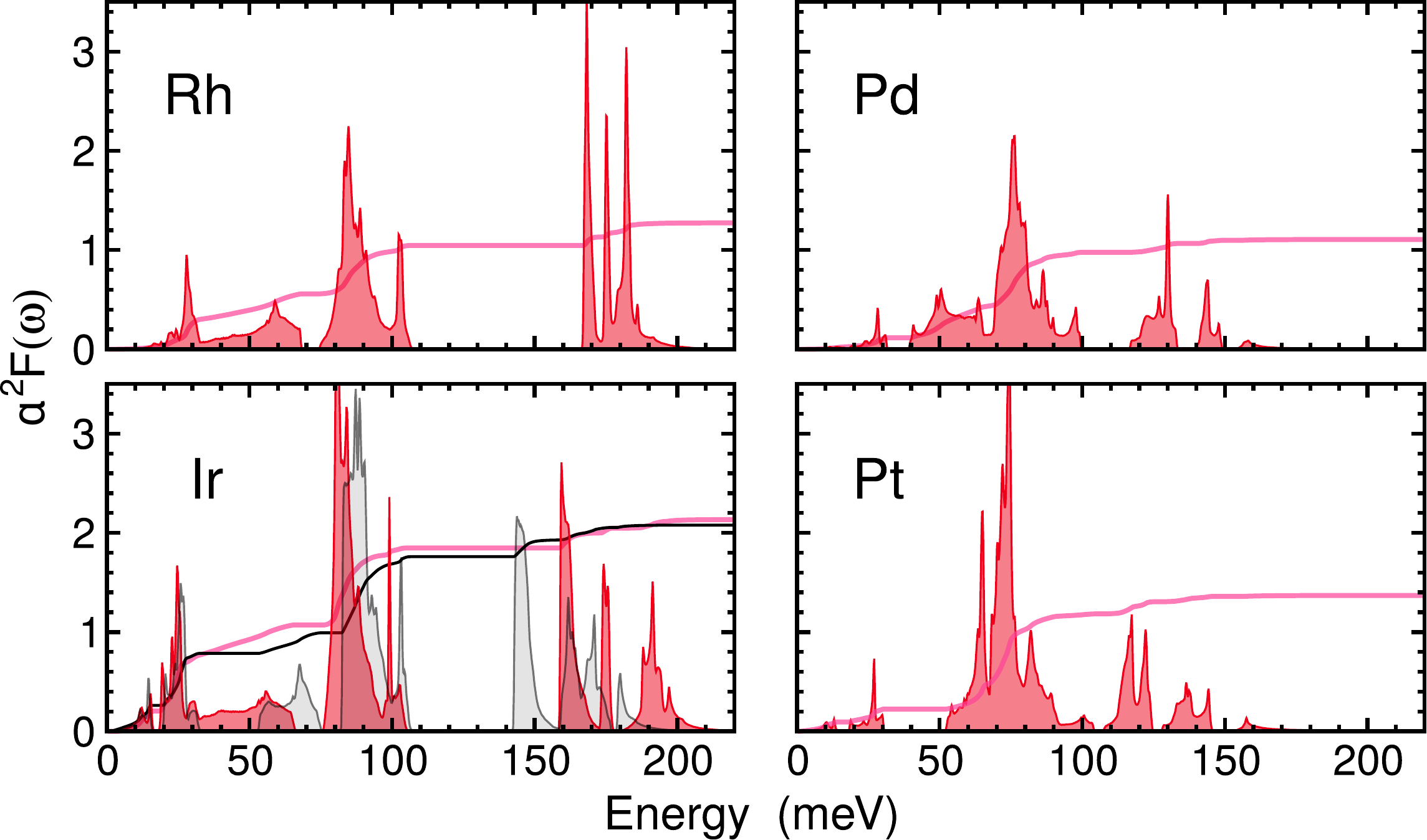}
  \caption{Eliashberg spectral function $\alpha^2F(\omega)$ for the four compounds studied here.
  The gray curve in the Ir panel is the Eliashberg phonon spectral function as simulated using the anharmonic phonon frequencies obtained with the SSCHA method. 
  The integration lines correspond to the $\lambda$ integration. 
  } 
  \label{fig:a2F}
\end{figure}

The electron phonon coupling is dominated by the low and middle part of the phonon spectrum (see \cref{fig:a2F}) implying that in all four materials at least 2/3 of the integrated coupling constant $\lambda$ is provided by hydrogen modes. The characteristic phonon frequency \olog\ ranges from 50~meV in the Ir compound to 66~meV in the Rh compound. These values are about half those in high pressure hydrides such as \ce{LaH10} or \ce{SH3}~\cite{Ma_SH3_JChemPhys2014}, nevertheless these phonon energies are quite high for room pressure superconductors.

\begin{table}
\caption{Electron-phonon coupling constant $\lambda$, logarithmic average of the phonon frequencies \olog\ (in meV), density of states at the Fermi level \dosef\ (in eV$^{-1}$), transition temperatures \Tc\ (in K) and superconducting gaps $\Delta$ (in meV) for the four compounds studied here. Calculations were performed with the isotropic Eliashberg equations and with SCDFT (both isotropic and anisotropic).}
\label{tab:super}
\begin{center}
\begin{tabular}{ l@{\hskip 0.5cm}ccc@{\hskip 0.5cm}cc@{\hskip 0.5cm}cc@{\hskip 0.5cm}c }
     &   &  &  & \multicolumn{2}{c}{Eliashberg} & \multicolumn{2}{c}{SCDFT} &  SCDFT \\
     &  &  &  & \multicolumn{2}{c}{isotropic}  & \multicolumn{2}{c}{isotropic} &  anisotropic \\
     & $\lambda$ & \olog & \dosef & \Tc        &  $\Delta$             &   \Tc        &  $\Delta$        &  \Tc  \\     \hline\\[-2mm]
 Rh  & 1.3 & 66 & 2.1 & 48.5    &  8.4                  &   45.1       &  8.1             &  52           \\
 Pd  & 1.1 & 65 & 0.9 & 66.5    & 11.2                  &   50.6       &  9.7             &  52           \\
 Ir  & 2.1 & 50 & 2.6 &77.0    & 16.3                  &   65.9       & 13.0             &  71           \\
 Pt  & 1.4 & 60 & 1.0 & 80.4    & 15.0                  &   64.1       & 12.5             &  66           \\
\end{tabular}
\end{center}
\end{table}

We have opted to compute the superconducting \Tc\ with three different methods: isotropic superconducting density-functional theory (SCDFT), anisotropic SCDFT, and isotropic Eliashberg. Usually the three approaches yield quite similar results, however our four materials are electronically quite peculiar. Note that the isotropic approximation is formally different in Eliashberg and in SCDFT as the former neglects the energy variation of the density of states in the phononic energy window. The spread of \Tc\ values predicted by different methods helps us to associate a theoretical errorbar to our predictions.

\begin{figure}[htb]
  \centering
  \includegraphics[width=\columnwidth]{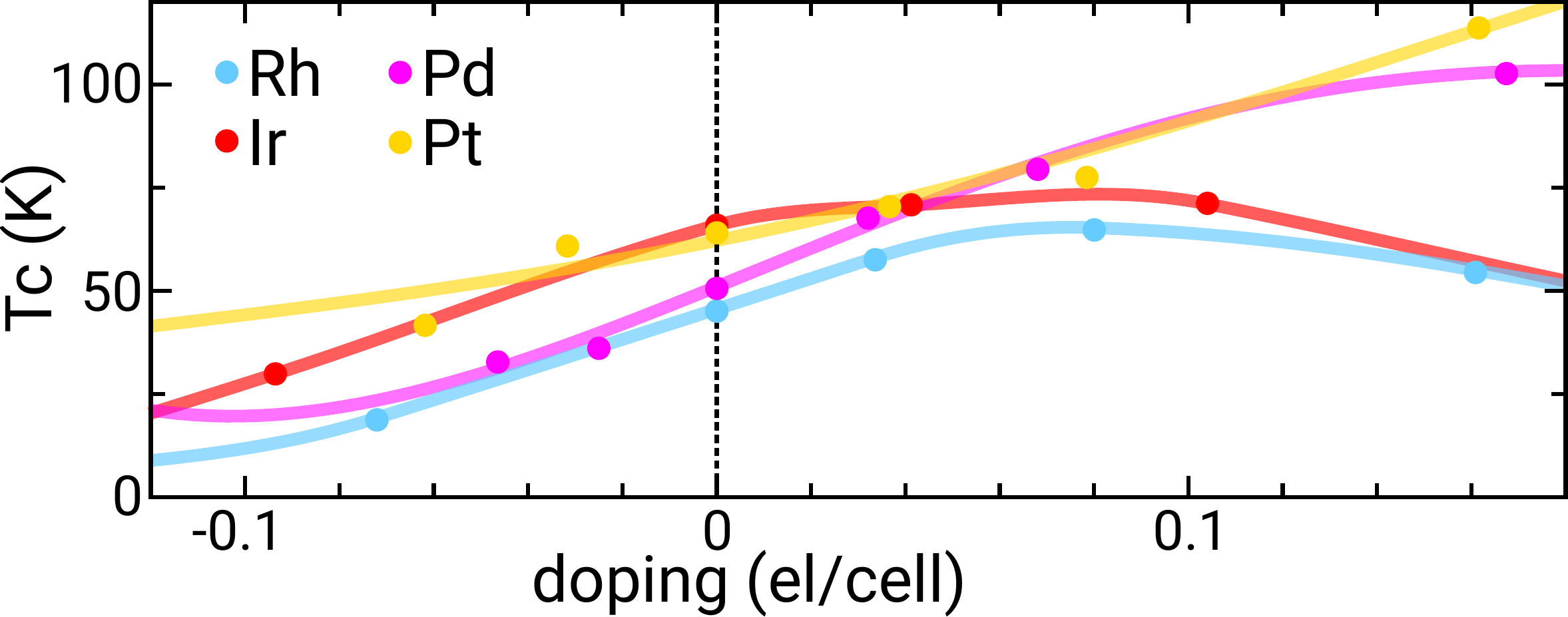} 
  \caption{Evolution of the superconducting transition temperature as a function of doping. The calculations were performed by moving rigidly the Fermi level and using the isotropic SCDFT approach.}
  \label{fig:doping}
\end{figure}

Results are collected in \cref{tab:super}. Our predictions for \Tc\ can be summarized in a range 45--50~K for the Rh compound, 51--67~K for the Pd compound, 66--77~K for the Ir compound and 64--80~K for the Pt compound. Naturally, such high values are due to the relatively high values of both \olog\ and $\lambda$ in these compounds. We believe that the lower \Tc\ in SCDFT is probably due to some limitations of the present functional at very high coupling. As shown in fig. \ref{fig:a2F}, anharmonicity barely impacts the value of $\lambda$ and $\olog$ for the Ir compound, leaving the prediction of \Tc\ unaffected. A similar minor impact is expected as well for the rest of the studied systems.

The complex electronic structure has further implications on the superconducting properties. A strongly varying density of states at the Fermi level means that these systems are expected to be particularly sensitive to doping. The Rh and Ir compounds, having a sharp peak at the Fermi level, are expected to be negatively affected both by electron doping and by hole doping, so that  \Tc\ in both cases might end up being smaller than that for the pristine materials. Calculations indicate that the maximal \Tc\ is achieved for a slight electron doping, of the order of 0.08 extra electrons per cell. In Pt and Pd compounds even a small hole doping completely destroys superconductivity, as the density of states drops abruptly below the Fermi level.  On the other hand, electron doping is expected to be beneficial, increasing \Tc. A simulation of the effect of doping on \Tc\ is reported in \cref{fig:doping}. Note, however, that these simulations are performed with a rigid shift approach, and therefore do not account for the disorder and structural deformations which are associated with doping, especially if this is induced by structural defects and impurities. 

Finally, we would like to discuss what happens when Mg is replaced by either Be or Ca. In the former case, all systems are very destabilized in the thermodynamical sense, and can be found at distances to the convex hull that exceed 200~meV/atom. Furthermore, it turns out that all systems are dynamically unstable, with highly imaginary phonons already at $\Gamma$ in the harmonic approximation.
The case with Ca is more interesting. We find that \ce{Ca2\{Rh,Ir\}H6}, are all stable, while \ce{Ca2PdH6} is 41~meV/atom and \ce{Ca2PtH6} is 30~meV/atom above the convex hull of thermodynamically stability. All these compounds are also dynamically stable, however \ce{Ca2\{Rh,Ir\}H6} are not superconducting, while \ce{Ca2\{Pd,Pt\}H6} are superconducting but with a transition temperature more than four times smaller than their Mg counterparts. Taking \ce{\{Mg,Ca\}2IrH6} as an example, the major difference between them is the lattice constant, that increases by 8\% in the Ca compound, while the X--H distance remains virtually unaltered. We then observe a noticeable change in the dispersion of the electronic valence band, leading to a considerable decrease in the \dosef\ by 20\%. The acoustic and the lowest six optical phonon bands, composed mostly of Ir and Mg/Ca character, are very similar in both Mg and Ca compounds. However the higher lying optical branches, composed of H vibrations, are much less dispersive for the Ca compound, which reflects a much smaller coupling with the electrons.

\section{Conclusion}

In conclusion, we propose a family of hydrides of the form \ce{Mg2XH6} as superconductors with \Tc\ reaching liquid nitrogen temperatures. These systems are constituted by isolated \ce{XH6} octahedra, with the large coordination of the metal X atoms made possible by the noble metal. They are furthermore thermodynamically stable (or close to it), and are therefore excellent candidates for experimental synthesis and characterization. The predicted superconducting hydrides could be efficiently identified by high-throughput synthesis and characterization of thin film materials libraries. Magnesium seems to be essential for this family as substitution by the isoelectronic Be or Ca leads to dynamically unstable systems (for the former) or to a very much reduced superconducting transition temperature (for the latter).

Conventional superconductivity at high-temperatures is enabled by the particular electronic structure, that includes a sizeable amount of hydrogen states at the Fermi level that couple strongly with the high-energy vibrations of the hydrogen atoms. In spite of this strong coupling, the systems remain dynamically stable. Anharmonic effects lead to a small change of the phonon frequencies and, consequently, do not impact significantly neither the physics of the systems nor the superconducting properties. It is also important to point out that the materials we have identified are characterized by the presence of sharp van Hove singularities at or near the Fermi level. This makes all these compounds quite sensitive to charge doping indicating that, unlike typical phononic superconductors, experimental synthesis might require extreme attention to sample quality.

While the occurrence of high-temperature superconductivity in hydrides at ambient pressure remains very rare, our findings indicate that specific systems hold the potential to meet the stringent criteria of thermodynamic stability, dynamical stability, and strong coupling to vibrations, opening exciting possibilities for further research and development of superconducting materials.

\section{Methods}
\label{sec:methods}

\subsection{Structure prediction and geometry relaxation}

We performed geometry optimizations and total energy calculations with the code VASP~\cite{Kresse1996,Kresse1996_1}. We applied the projector augmented wave parameters~\cite{paw,paw2} of VASP version 5.2 with a cutoff of 520~eV. Geometries were converged until forces were smaller than 0.005~eV/\AA. As exchange-correlation functional we used the Perdew-Burke-Ernzerhof for solids~\cite{PBEsol}. The Brillouin zones were sampled by uniform $\Gamma$-centered $k$-point grids with a density of 2000 and 8000 $k$-points per reciprocal atom for the geometry optimization and the final energy evaluation respectively. 

For the global structure prediction we used the Minima Hopping Method (MHM)~\cite{Goedecker2004,Amsler2010}. We performed MHM runs for all the stoichiometries within the Mg$_x$Ir$_y$H$_z$ formula, with $1 \leq x,y,z \leq 6$ and $x+y+z \leq 10$. This amounts to a total of 98 unique chemical compositions. For a given stoichiometry, the initial geometries were obtained randomly, ensuring only that the minimal distance between the atoms was at least equal to the sum of the covalent radii. Each minima hopping run was repeated twice. The 12 lowest energy structures for each composition were then re-optimized.

\subsection{Phonons and electron-phonon}
Harmonic phonon frequencies and electron phonon matrix elements are computed on a $16\times16\times16$ $k$-grid and a $10\times10\times10$ $q$-grid within density functional perturbation theory~\cite{PhysRevLett.58.1861,Baroni_DFPT_RMP2001} (DFPT) as implemented in the \textsc{quantum espresso} package~\cite{espresso1,espresso2}. Electronic occupation numbers were set by a Methfessel-Paxton function of 0.02 Ry width~\cite{MethfesselPaxton_PRB1989}. We used pseudopotentials from the \textsc{pseudodojo} project~\cite{vanSetten2018pseudodojo}, specifically the \textit{stringent} norm-conserving set. The plane-wave expansion cutoff was set at 120~Ry.

\subsection{SSCHA calculations}

Due to the low mass of H, we used the SSCHA method to study quantum ionic and anharmonic effects~\cite{Errea-PRB2014-SSCHA,Bianco-PRB2017-SSCHA,monacelli2018pressure,Monacelli2021}. SSCHA calculations were performed with a 2$\times$2$\times$2 supercell including 72 atoms, which yields the dynamical matrices on a commensurate 2$\times$2$\times$2 \textbf{q}-mesh. The plane-wave expansion cutoff was set at 96~Ry and a $4\times4\times4$ $k$-grid in the supercell was sufficient to converge the SSCHA gradient.  The phonon frequencies obtained with the SSCHA in the 2$\times$2$\times$2 \textbf{q}-mesh were transferred to the 10$\times$10$\times$10 \textbf{q}-grid used for the superconductivity calculations by interpolating the difference between the SSCHA and harmonic dynamical matrices. The SSCHA phonon frequencies and polarization vectors used in superconductivity calculations were obtained from the Hessian of the SSCHA free energy and not relaxing the unit cell within the SSCHA. However, the $x$ free parameter associated to the H atoms in the 24e positions was relaxed within the SSCHA, but only varied by a 0.7\%, not affecting the structure. The SSCHA was also repeated relaxing also the cell parameters, but the impact was minor. 

\subsection{Superconductivity}

Superconductivity simulations were performed with three different methods. Isotropic SCDFT, anisotropic SCDFT and isotropic Eliashberg theory.
In the isotropic SCDFT calculations we neglected anisotropy (i.e. $k$-dependence) of the electron-phonon matrix elements by assuming an homogeneous coupling for all electronic states. The isotropic coupling was computed strictly at the Fermi level (with an integration window of 90~meV), and described by the \aF\ function. The energy dependence of the density of electronic states was taken into account both in computing the diagonal and the off-diagonal parts of the SCDFT phononic kernels. Usually the isotropic approximation gives results which are close to the fully anisotropic case unless the material has a clear multigap structure. Anisotropic calculations give slightly larger values of \Tc, from a few percent to about 20\% larger values. 
Isotropic Eliashberg predicts critical temperatures which are between 10 and 20\% larger than the isotropic SCDFT predictions. Formally neither theory is superior to the other. On one hand Eliashberg theory only relies on a numerical Matsubara summation of exchange diagrams, without further approximations. However owing to the computational cost phononic contributions are strictly evaluated at the Fermi level, and the effect of a fast varying Fermi density of states is neglected. On the other hand in SCDFT the energy integration is fully performed, however the Matsubara summation is not done explicitly but hidden in the exchange-correleation functional of the theory, which hides some additional approximations as compared to a direct evaluation of the diagrams.
All SCDFT~\cite{OGK_SCDFT_PRL1988,Lueders_SCDFT_PRB2005,Marques_SCDFT_PRB2005} simulations were performed with the exchange correlation phononic functional from Ref.~\cite{SPG_EliashbergSCDFT_PRL2020}. Eliashberg simulations were done including ab initio Coulomb  interactions with the approach proposed in Refs.~\cite{Adavydov_PRB2020,Sanna_AnisoCaC6_PRB2007} and implemented with the simplified scheme described in Ref.~\cite{Pellegrini_SimpEliashberg_JOPM2022}. 
The RPA Coulomb interactions were computed on a $8\times8\times8$ $k$-grid in the adiabatic LDA approximation~\cite{PerdewZunger_LDA_1981} as implemented in the elk code~\cite{ElkCode}.

\section{Data availability statement}

All relevant data is present in the manuscript and in the Supplementary
Information.

\section{Acknowledgements}
T.F.T.C acknowledges financial support from FCT - Fundação para a Ciência e Tecnologia, Portugal (projects UIDB/04564/2020 and 2022.09975.PTDC) and the Laboratory for Advanced Computing at University of Coimbra for providing HPC resources that have contributed to the research results reported within this paper. M.A.L.M. acknowledges partial funding from Horizon Europe MSCA Doctoral network grant n.101073486, EUSpecLab, funded by the European Union, and from the Keele Foundation. 
Y.-W.F. and I.E. received funding from the European Research Council (ERC) under the European Union’s Horizon 2020 research and innovation programme (Grant Agreement No. 802533) and acknowledge PRACE for awarding access to the EuroHPC supercomputer LUMI located in CSC's data center in Kajaani, Finland through EuroHPC Joint Undertaking (EHPC-REG-2022R03-090). I.E. also acknowledges funding from the Spanish Ministry of Science and Innovation (Grant No. PID2022-142861NA-I00) and the Department of Education, Universities and Research of the Basque Government and the University of the Basque Country (Grant No. IT1527-22).

\section{Author Contributions}

T.F.T.C. and M.A.L.M. performed the machine learning prediction and the preliminary analysis of the superconducting properties. A.S. performed the electron-phonon and the SCDFT calculations. Y.-W.F. and I.E. performed the calculations of quantum anharmonic effects on the phonons. All authors contributed to designing the research, interpreting the results and writing of the manuscript.

\section{Competing  Interests}

The authors declare that they have no competing interests.

\end{document}